\documentstyle[epsbox,sf99proc]{article}
\def\Msun{M_{\odot \hskip-4.8pt \bullet}}

\title{%
NMA Observations of CO(2-1) and CO(1-0) Emission in the Starburst
Region of NGC 4527} 
\authors{
Shibatsuka, T.\affilmark{1},
Matsushita, S.\affilmark{2},
Kohno, K.\affilmark{3},
Kawabe, R.\affilmark{3}
}

\affiltexts{
\affiltext{1}{University of Tokyo, Japan},
\affiltext{2}{Graduate University for Advanced Studies, Japan},
\affiltext{3}{Nobeyama Radio Observatory, Japan}
}

\firstauthor
{TS}
{shiba@nro.nao.ac.jp}

\authorsADS{%
Shibatsuka, T.;
Kohno, K.;
Matsushita, S.;
Kawabe, R.;
}

\affiliationsADS{%
AA(University of Tokyo)
AB(Nobeyama Radio Observatory)
AC(Graduate University for Advanced Studies)
AD(Nobeyama Radio Observatory)
}

\begin{document}

%
\begin{abstract}
We have performed high resolution CO(2-1), CO(1-0), HCN(1-0) and
 HCO$^+$(1-0) observations of a ``low star formation efficiency
 starburst galaxy'' NGC 4527 with the Nobeyama Millimeter Array. The
 integrated intensity ratios, CO(2-1)/CO(1-0) and HCN(1-0)/CO(1-0), are
 found to be
 0.6$\pm0.05$ and 0.06$\pm0.007$, respectively, at the center. These line
 ratios are smaller 
 than those in prototypical starburst galaxies such as NGC 253 and M82,
 and we suggest that the fraction of dense molecular gas to the total
 molecular mass in the central a few kpc region of NGC 4527 is
 small. This fact may be responsible for the low star formation
 efficiency in the center of NGC 4527.
\end{abstract}

\section{Introduction}
NGC 4527 is a nearby spiral galaxy (Table 1), and has been classified as
a starburst galaxy because of it's abundant molecular gas ($M_{\rm H_2}
\sim 4 \times 10^9 \Msun$) and high {\it star formation rate} (Young \&
Devereux 1991). However, Young \& Devereux (1991) pointed
out that the star formation rate per unit gas mass indicated by
$L_{\rm IR}/M_{\rm H_2}$, the  {\it star formation efficiency}, of NGC
4527 is very low compared 
with their starburst sample. What controls the star formation
properties in the central region of NGC 4527?
In order to address the relationship between the physical conditions of
molecular gas and star formation, we have performed high resolution 
multiple molecular emission observations in the central a
few kpc region of NGC 4527 with the Nobeyama Millimeter Array (NMA).

%
\begin{table}[ht]
\begin{center}
\begin{tabular}{lll} \hline \hline
    Parameter  &  Value     & Ref.   \\ \hline
    Morphology &  SAB(s)bc  & RC3    \\
               &  Sb(s)II   & RSA    \\
    Position of nucleus &&    (1)    \\
    $\alpha$(B1950)&12$^{\rm h}$31$^{\rm m}$35$^{\rm
    s}\hspace{-3.8pt}.\hspace{1.2pt}$1&\\ 
    $\delta$(B1950)&+02$^{\circ}$55$'$47$.\hspace{-2pt}''$0&\\ 
    Position angle & 64$^{\circ}$&(2)\\
    Inclination angle & 72$^{\circ}$&(3)\\
    Adopted Distance & 13.5 Mpc&(4)\\
    Linear scale & 65 pc arcsec$^{-1}$&\\
    $I_{\rm CO}$   &15.8 K km s$^{-1}$ &(5)\\
    $S_{\rm CO}$   &658 $\pm$ 50 Jy km s$^{-1}$ &(6)\\
 \hline \hline
\end{tabular}
\caption{Properties of NGC 4527. Reference (1)(2)Hummel et al. 1987; (3)
 Rubin et al. 1997; (4) Tully 1988; (5)(6) Young \& Devereux 1991;}
\end{center}
\end{table}
\vspace{-2mm}

\section{Observations}

The central region of NGC 4527 was observed in 
multiple line with the NMA which consists of six 10 m
dishes. The observations were made during December 1998 to May 1999. The 
Ultra Wide-Band Correlator(UWBC) with a bandwidth of 1024 MHz enables us
to perform simultaneous observations of both HCN(1-0) and HCO$^+$(1-0).
The uncertainty in the absolute flux scale is estimated to be about $\pm10$\%.

%
\begin{figure}[ht]
\begin{center}
\psbox[width=80mm,vscale=1.0]{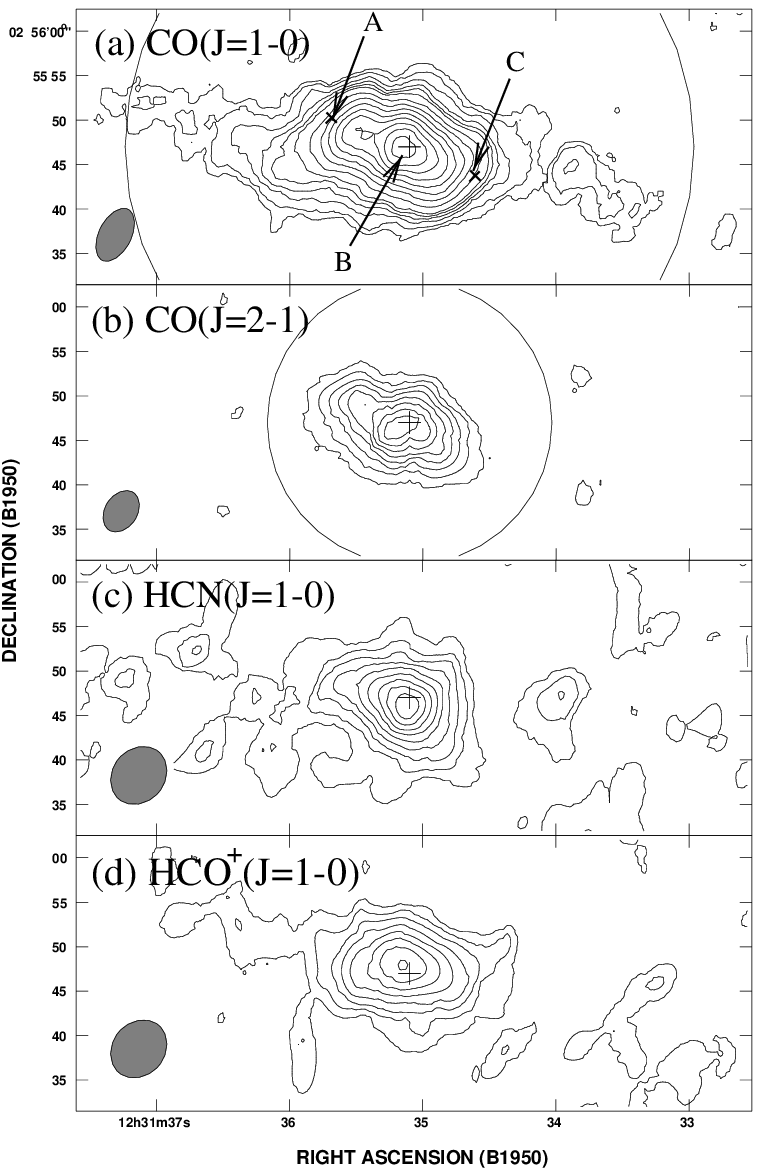}
\end{center}
\caption{(a)Total intensity distribution of CO(1-0). Contour levels are 
1, 2, 4, 6, 8, 10, 15, 20, 25, 30, 35, 40, 45 and 50 $\sigma$, where 1
 $\sigma$ = 1.82 Jy beam$^{-1}$ km s$^{-1}$. A cross (+) marks the
 field center and radio continuum peak. A$\sim$C indicate the positions where
intensity ratios were calculated in Table 3.
 (b)Total intensity  distribution of CO(2-1). Contour levels are
 2, 4, 6, 8, 10 and 12 $\sigma$, where 1 $\sigma$ = 12 Jy beam$^{-1}$ km
 s$^{-1}$. (c)(d)Total intensity distribution of HCN(1-0) and
 HCO$^+$(1-0), respectively. Contour levels are 1, 2, 3, 4, 5,
 6, 7, 8 and 9 $\sigma$, where 1 $\sigma$ = 0.756 Jy beam$^{-1}$ km
 s$^{-1}$.}   
\label{fig:1}
\end{figure}

%
\begin{table*}[ht]
\begin{center}
\begin{tabular}{cccccc} \hline \hline
Line and transition  &  CO(2-1)    & CO(1-0) & CO(1-0) & HCN(1-0)  &
 HCO$^+$(1-0) \\
&&&(UV tapered 40 k$\lambda$)&& \\ \hline 
Array& D&AB, C, D&AB, C, D&C, D&C, D\\
Synthesized beam & $4^{\prime \prime}.7 \times 3^{\prime \prime}.6$&
$4^{\prime \prime}.0 \times 2^{\prime \prime}.0$ &
$6^{\prime \prime}.0 \times 3^{\prime \prime}.5$&
\multicolumn{2}{c}{$7^{\prime \prime}.0 \times 6^{\prime \prime}.0$}\\
 &($300 \times 230$ pc)&($260 \times 130$ pc)&($390 \times 230$
 pc)&\multicolumn{2}{c}{($460 \times 390$ pc)}\\
Peak integrated intensity&330 K km s$^{-1}$&600 K km s$^{-1}$&420 K km
 s$^{-1}$&26 K km s$^{-1}$&23 K km s$^{-1}$\\ 
Total integrated flux&1020 Jy km s$^{-1}$&540 Jy km s$^{-1}$&560 Jy km
 s$^{-1}$&17 Jy km s$^{-1}$&17 Jy km s$^{-1}$\\ 
 \hline \hline
 \end{tabular}
\caption{NMA observations.}
\end{center}
\end{table*}

%
\begin{table*}[ht]
\begin{center}
\begin{tabular}{llllllll} \hline \hline
    Galaxy  &  Star formation or Position & $R_{\rm 2-1/1-0}$ &
Ref.& $R_{\rm HCN/CO}$ &   Ref. &$R_{\rm HCO^+/HCN}$ & Ref. \\ \hline
 NGC 4527 & A (r = 400 pc) & 0.48   &(1)& 0.052&(1)&1.0&(1)  \\
          & B (r $<$ 200 pc) & 0.58  &(1)& 0.060&(1)&0.9&(1)  \\
          & C (r = 400 pc) & 0.48   &(1)& 0.060&(1)&1.0&(1)  \\ \hline 
 NGC 253  & nuclear starburst & 1.1 &(2)& 0.3&(6)&0.9&(10) \\
 M82      & nuclear starburst & 1.3 &(3)& 0.2&(7)&2.0&(10) \\
 IC 342   & moderate starburst& 1.1 &(4)& 0.16&(8)&0.5&(10) \\ 
 Milky Way& recent burst of star formation  & 0.65&(5)&
 0.08&(9)&$\cdots$&$\cdots$ \\ 
 \hline \hline
 \end{tabular}
\caption{Integrated intensity ratios of NGC 4527 and other galaxies.
 A$\sim$C indicate the positions where intensity ratios were calculated in
 Figure 1a. Reference. (1)this work; (2)Aalto et al. 1995; (3)Wild et al.
 1992; (4)Eckart et al 1990; (5)Oka et al. 1996;  (6)Sorai
 1997; (7)Shen \&  Lo 1995; 
 (8)Downes et  al. 1992; (9)Jackson et al. 1996;
 (10)Rieu et  al. 1992; }
\end{center}
\end{table*}

\section{Results}
\subsection{CO(1-0)}

The CO(1-0) map in Figure 1 shows two outstanding features. One is
the strong concentration of CO emission toward the nucleus (r $<$ 5$''$
 or 350 pc) with a slight elongation along the major axis of the
galaxy (P.A. = 64$^\circ$). Another feature is two offset ridges of CO
emission extends over r $<$ 27$''$ (1800 pc) along P.A. of
64$^\circ$. The overall distribution of
the CO in NGC 4527 is very similar to that of barred spiral galaxies,
which shows two offset ridges at the leading edges of a bar with a central
concentration of gas (e.g., Kenney et al. 1992), although the optical
morphology can not make clear whether the presence of the bar (Table 1).
Our CO(1-0) map recovers about 80 \% of a single dish flux (Young et
al. 1995). Molecular gas mass estimated from the
observed CO flux is about $1.2 \times 10^9 $ $\Msun$, and the peak
molecular gas surface density, $\Sigma _{\rm H_2}$, at the center is $8.8
\times 10^2 $ $\Msun$ pc$^{-2}$ adopting a Galactic CO-to-H$_2$ conversion 
factor ($3 \times 10^{20} $cm$^{-2} {\rm (K\ km\ s^{-1})}^{-1}$). This
velocity width (full width of zero intensity) of 500 km s$^{-1}$ is the
same as that of the single-dish line profiles obtained with the FCRAO 14 
m (Young et al. 1995) and NRAO 12 m (Helfer \& Blitz 1993).

\subsection{CO(2-1)}

The distribution of the CO(2-1) emission resembles the CO(1-0) map,
although the extended ridges seen in CO(1-0) map (r $<$ 27$''$ or 1800
pc) do not appear in this CO(2-1) map. It is probably
due to insufficient sensitivity of CO(2-1) observations, and also the
lack of short spacing. The summary of CO observations are listed in 
Table 2.

\subsection{HCN(1-0) and HCO$^+$(1-0)}
HCN(1-0) and HCO$^+$(1-0) emission concentrated toward the center have
been detected. Extended components correspond to the CO ridges may be also
seen in the maps, although the S/N ratios are insufficient to make
detail comparisons.  
We assumed almost all the single dish flux is recovered in both HCN and
HCO$^+$ 
maps, and the previous observations of HCN emission seems to support the
validity of this assumption (see Helfer \& Blitz 1997 and
Kohno et al. 1999a).  We summarize HCN and HCO$^+$ observations in Table
2.

\section{Discussion}
\subsection{Integrated intensity ratios}

The CO(2-1) to CO(1-0) integrated intensity ratio, $R_{\rm 2-1/1-0}$, is
a measure of dense gas fraction to the total 
molecular gas (e.g., Sakamoto et al. 1994), and sensitive in a density
range of about 10$^2$ - 10$^4$ cm$^{-3}$. We convolved the CO(1-0) map
to the same beam size as that of the CO(2-1) to calculate $R_{\rm
2-1/1-0}$ values. The peak $R_{\rm 2-1/1-0}$
of about 0.6 at the center of NGC 4527 is somewhat smaller than 
$R_{\rm 2-1/1-0}$ values($\sim 0.89$) in the
central regions of spiral galaxies (Braine \& Combes 1992), suggesting a
presence 
of a large amount of low density molecular gas in this region. Note the
observed $R_{\rm 2-1/1-0}$ could contain considerable error due to lack
of single dish CO(2-1) measurements, however. A radial decrease of
$R_{\rm 2-1/1-0}$ may be apparent; the ratio is about 0.6 at the 
center, while the ratio decreases to 0.5 at r = 6$''$ (400
pc).

The HCN(1-0) to CO(1-0) integrated intensity ratio, $R_{\rm HCN/CO}$, is
another measure of dense gas fraction to the total molecular 
gas (e.g., Kohno et al. 1999), and sensitive in a density
range of about 10$^3$ - 10$^5$ cm$^{-3}$. The observed peak
$R_{\rm HCN/CO}$ of 0.06 is significantly smaller than a 
$R_{\rm HCN/CO}$ values in starburst galaxies such as NGC 253 (0.2 $\sim
0.3$; e.g., Paglione et al. 1997; Sorai 1997) and M82 ($\sim 0.2$; Shen
\& Lo 1995), 
and is similar to those in ``normal'' galaxies such as the Galactic 
Center ($R_{\rm HCN/CO} \sim 0.08$; Paglione et al. 1998).

\subsection{Dense molecular gas and star formation}

Because stars are formed from dense cores of molecular clouds rather
than their diffuse envelopes(e.g. Lada 1992), study of dense molecular
gas is essential to understand star formation in galaxies. In fact, good 
spatial coincidence between HCN and H$\alpha$ emission has been
reported in nearby starburst galaxies (e.g. Paglione et al. 1997; Kohno
et al 1999).

Both $R_{\rm 2-1/1-0}$ and $R_{\rm HCN/CO}$ suggest that the fraction of
dense molecular clouds to the total molecular gas in the center of NGC
4527 is smaller than that of prototypical starburst galaxies such as NGC 
253 and M82.

It has been demonstrated that the fraction of dense molecular gas
measured with the $R_{\rm HCN/CO}$ correlates with the star formation
efficiency (Solomon et al. 1992), which
is indicated by  $L_{\rm IR}/L_{\rm CO}$. 

Considering these results, we suggest that the small dense gas fraction
is responsible for the low star formation efficiency in the center of
NGC 4527. But the star formation rate of NGC 4527 would be enhanced existence
of rich gas in central region.

%
\section*{References} \vspace{1mm}

\reference
Aalto et al. 1995, A\&A, 300, 369

\reference
Braine \& Combes 1993, A\&AS, 97, 887

\reference
Casoli \& Gerin 1993, A\&A, 279, L41

\reference
de Vaucouleurs et al. 1991, Third Reference Catalogue of Bright Glaxies
(New York: Springer-Verlag)

\reference
Downes et al. 1992, A\&A, 262, 424

\reference
Eckart et al 1990, ApJ, 348, 434

\reference
Gerin et al. 1991, A\&A, 251, 32

\reference
Helfer \& Blitz 1993, ApJ, 419, 86

\reference
Helfer \& Blitz 1997, ApJ, 478, 162

\reference
Hummel et al. 1987, A\&AS, 70, 517

\reference
Jackson et al. 1996, ApJ, 456, L91

\reference
Kenney et al. 1992, ApJ, 395, L79

\reference
Kohno et al. 1999, ApJ, 511, 157

\reference
Lada 1992, ApJ, 393, L25

\reference
Oka et al. 1996, ApJ, 460, 334

\reference
Paglione et al. 1997, ApJ, 484, 656

\reference
Paglione et al. 1998, ApJ, 493, 680

\reference
Rieu at al. 1992, ApJ, 399, 521

\reference
Rubin et al. 1997, A.J, 113, 1250

\reference
Sakamoto et al. 1994, ApJ, 425, 641

\reference
Solomon et al. 1992, ApJ, 387, L55

\reference
Sorai 1997, PhD thesis, University of Tokyo

\reference
Shen \& Lo 1995, ApJ, 334, L99

\reference
Tully 1998, Nearby Galaxies Catalog (Cambridge: Cambridge Univ. Press)

\reference
Wild et al. 1992, A\&A, 265, 447

\reference
Young \& Devereux 1991, ApJ, 373, 414

\reference
Young et al. 1995, ApJS, 98, 219

\vspace{-1mm} 

\end{document}